\documentclass[12pt]{article}
\usepackage{pic02}
\usepackage{hyperref}
\usepackage{url}
\usepackage{graphicx}

\newcommand{\rmt}{\rm\textstyle}
\newcommand{\rms}{\rm\scriptstyle}
\newcommand{\nub}{\overline{\nu}}
\newcommand{\nubar}[0]{\overline{\nu}}

\newcommand{\nunub}{\stackrel{{\footnotesize (-)}}{\nu}}
\newcommand{\txnunub}[1]{\nu_{#1},\nub_{#1}}

\newcommand{\stw}{\mbox{$\sin^2\theta_W$}}
\newcommand{\stwos}{\mbox{$\sin^2\theta_W^{\rms(on-shell)}$}}

\newcommand{\Rnu}{\mbox{$R^{\nu}$}}
\newcommand{\Rnub}{\mbox{$R^{\nub}$}}

\newcommand{\qbar}[0]{\overline{q}}

\newcommand{\sbar}[0]{\overline{s}}

\newcommand{\Ubar}[0]{\overline{U}}

\newcommand{\Sbar}[0]{\overline{S}}
\newcommand{\Dbar}[0]{\overline{D}}
\newcommand{\uubar}[0]{\mbox{$\stackrel{(-)}{u}$}}
\newcommand{\ddbar}[0]{\mbox{$\stackrel{(-)}{d}$}}

\newcommand{\savbar}[0]{\langle\overline{s}(x)\rangle}

\newcommand{\sav}[0]{\langle{s(x)}\rangle}

\newcommand{\ARCHIVE}{1}  

\begin{document}

\title{\bf OFF THE MASS SHELL: ELECTROWEAK PHYSICS AT NUTEV}
\author{The NuTeV Collaboration \footnotemark, 
represented by\\
Kevin~S.~McFarland\\ 
{\em University of Rochester, Dept.\ of Physics and Astronomy, Rochester, NY 14627 USA}
}
\maketitle

%
%
\begin{figure}[h]
\begin{center}
%
%
{\ifnum\ARCHIVE=0
\includegraphics[height=4.5cm]{kevin.ps}
%
%
\else
\vspace{4.5cm}
\fi}
\end{center}
\end{figure}

\baselineskip=14.5pt
\begin{abstract}
  The NuTeV collaboration
  has performed precision measurements of the ratio of
  neutral current to charged current cross-sections in high rate, high energy
  neutrino and anti-neutrino beams on a dense, primarily steel, target.  The
  separate neutrino and anti-neutrino beams, high statistics, and improved
  control of other experimental systematics, allow the determination of
  electroweak parameters with significantly greater precision than past $\nu
  N$ scattering experiments.  Our null hypothesis test of the standard model
  prediction measures $\stwos=0.2277\pm0.0013({\rmt stat})\pm0.0009({\rmt
    syst})$, a value which is $3.0\sigma$ above the prediction.  We discuss 
  possible explanations for and implications of this discrepancy.
\end{abstract}
\newpage

\baselineskip=17pt

\section{Introduction and Motivation}

{\def\thefootnote{*}  
\footnotetext{ The NuTeV Collaboration:
T.~Adams$^{4}$, A.~Alton$^{4}$, S.~Avvakumov$^{8}$, 
 L.~de~Barbaro$^{5}$, P.~de~Barbaro$^{8}$, R.~H.~Bernstein$^{3}$, 
 A.~Bodek$^{8}$, T.~Bolton$^{4}$, J.~Brau$^{6}$, D.~Buchholz$^{5}$, 
 H.~Budd$^{8}$, L.~Bugel$^{3}$, J.~Conrad$^{2}$, R.~B.~Drucker$^{6}$, 
 B.~T.~Fleming$^{2}$, R.~Frey$^{6}$, J.A.~Formaggio$^{2}$, J.~Goldman$^{4}$, 
 M.~Goncharov$^{4}$, D.~A.~Harris$^{8}$, R.~A.~Johnson$^{1}$, J.~H.~Kim$^{2}$,
 S.~Koutsoliotas$^{2}$, M.~J.~Lamm$^{3}$, W.~Marsh$^{3}$, D.~Mason$^{6}$, 
 J.~McDonald$^{7}$, K.S.~McFarland$^{8}$, C.~McNulty$^{2}$, 
    D.~Naples$^{7}$, 
 P.~Nienaber$^{3}$, V.~Radescu$^{7}$, A.~Romosan$^{2}$, 
 W.~K.~Sakumoto$^{8}$, H.~Schellman$^{5}$,
 M.~H.~Shaevitz$^{2}$, P.~Spentzouris$^{2}$, E.~G.~Stern$^{2}$, 
 N.~Suwonjandee$^{1}$, M.~Tzanov$^{7}$, M.~Vakili$^{1}$, A.~Vaitaitis$^{2}$, 
 U.~K.~Yang$^{8}$, J.~Yu$^{3}$, 
 G.~P.~Zeller$^{5}$, and E.~D.~Zimmerman$^{2}$;
{
\em
$^1$University of Cincinnati, Cincinnati, OH 45221; 
$^2$Columbia University, New York, NY 10027;
$^3$Fermi National Accelerator Laboratory, Batavia, IL 60510;
$^4$Kansas State University, Manhattan, KS 66506;
$^5$Northwestern University, Evanston, IL 60208;
$^6$University of Oregon, Eugene, OR 97403;
$^7$University of Pittsburgh, Pittsburgh, PA 15260;
$^8$University of Rochester, Rochester, NY 14627
}
}}
Neutrino scattering played a key role in establishing the structure of the
standard model of electroweak unification, and it continues to be one of the
most precise probes of the weak neutral current available experimentally
today.  With copious data from the production and decay
of on-shell $Z$ and $W$ bosons for comparison, contemporary neutrino
scattering measurements serve to validate the theory over many orders of
magnitude in momentum transfer and provide one of the most precise tests of
the weak couplings of neutrinos.  In addition, precise measurements of weak
interactions far from the boson poles are inherently sensitive to processes
beyond our current knowledge, including possible contributions from
leptoquark and $Z^\prime$ exchange\cite{langacker} and new properties of
neutrinos themselves\cite{oscpaper}.

The Lagrangian for weak neutral current $\nu$--$q$
scattering can be written
as
\begin{eqnarray}
{\cal L}&=&-\frac{G_F\rho_0}{\sqrt{2}}(\nubar\gamma^\mu(1-\gamma^5)\nu)
        \nonumber\\
          &&\times\left( \epsilon^q_L {\qbar}\gamma_\mu(1-\gamma^5){q}+
                \epsilon^q_R {\qbar}\gamma_\mu(1+\gamma^5){q}\right) , 
\label{eqn:lagrangian}
\end{eqnarray}
where deviations from $\rho_0=1$ describe non-standard sources of SU(2)
breaking, and $\epsilon^q_{L,R}$ are the chiral quark couplings~\footnote{Note
  that although we use a process-independent notation here for a tree-level 
  $\rho$, radiative corrections to $\rho$ depend slightly
  on the particles involved in the weak neutral
  interaction.  In this case, $\rho\equiv \sqrt{\rho^{(\nu)}\rho^{(q)}}$.}
For the weak charged current, $\epsilon^q_L=I_{\rms weak}^{(3)}$ and
$\epsilon^q_R=0$, but for the neutral current $\epsilon^q_L$ and
$\epsilon^q_R$ each contain an additional term, $-Q\stw$, where $Q$ is the
quark's electric charge in units of $e$.

The ratio of
neutral current to charged current cross-sections 
for either $\nu$ or $\nub$ scattering
from isoscalar targets of $u$ and $d$ quarks can be written as\cite{llewellyn}
\begin{equation}
R^{\nu(\nub)} \equiv \frac{\sigma(\nunub N\rightarrow\nunub X)}
                 {\sigma(\nunub N\rightarrow\ell^{-(+)}X)}  
= (g_L^2+r^{(-1)}g_R^2),
\label{eqn:ls}
\end{equation}
where
\begin{equation}
r \equiv \frac{\sigma({\overline \nu}N\rightarrow\ell^+X)}
                {\sigma(\nu N\rightarrow\ell^-X)} \sim \frac{1}{2},  
\label{eqn:rdef} 
\end{equation}
and $g_{L,R}^2=(\epsilon^u_{L,R})^2+(\epsilon^d_{L,R})^2$.
Many corrections to Equation~\ref{eqn:ls} are required in a real
target\cite{nc-prl}, but those most uncertain result from the
suppression of the production of charm, which is the
CKM-favored final state for charged-current scattering from the strange sea.
One way to reduce this source of uncertainty on electroweak parameters 
is to measure the observable
 \begin{eqnarray}
R^{-} &\equiv& \frac{\sigma(\nu_{\mu}N\rightarrow\nu_{\mu}X)-
                   \sigma(\nub_{\mu}N\rightarrow\nub_{\mu}X)}
                  {\sigma(\nu_{\mu}N\rightarrow\mu^-X)-  
                   \sigma(\nub_{\mu}N\rightarrow\mu^+X)} \nonumber\\  
&=& \frac{\Rnu-r\Rnub}{1-r}=(g_L^2-g_R^2), 
\label{eqn:rminus}
\end{eqnarray}
first suggested by Paschos and Wolfenstein\cite{Paschos-Wolfenstein}
and valid under the assumption of equal momentum carried by the $u$
and $d$ valence quarks in the target.  Since $\sigma^{\nu
q}=\sigma^{\nub\, \qbar}$ and $\sigma^{\nub q}=\sigma^{\nu \qbar}$,
the effect of scattering from sea quarks, which are symmetric under
the exchange $q\leftrightarrow\qbar$, cancels in the difference of
neutrino and anti-neutrino cross-sections.  Therefore, the suppressed
scattering from the strange sea does not cause large uncertainties in
$R^-$.  $R^-$ is more difficult to measure than $R^\nu$, primarily
because the neutral current scatterings of $\nu$ and $\nub$ yield
identical observed final states which can only be distinguished
through {\em a priori} knowledge of the initial state neutrino.

The experimental details and theoretical treatment of cross-sections
in the NuTeV electroweak measurement are described in detail
elsewhere\cite{nc-prl}. In brief, we measure the experimental ratio of
neutral current to charged current candidates in both a neutrino and
anti-neutrino beam.  A Monte Carlo simulation is used to express these
experimental ratios in terms of fundamental electroweak parameters.
This procedure implicitly corrects for details of the
neutrino cross-sections and experimental backgrounds.  For the
measurement of $\stw$, the sensitivity arises in the $\nu$ beam,
and the measurement in the $\nubar$ beam is the control sample for
systematic uncertainties, as suggested in the Paschos-Wolfenstein
$R^-$ of Eqn.~\ref{eqn:rminus}.  For simultaneous fits to two
electroweak parameters, e.g., $\stw$ and $\rho$ or left and right
handed couplings, this redundant control of systematics cannot be
realized.

\section{Result}\label{sect:results}

As a test of the electroweak predictions for neutrino nucleon scattering,
NuTeV performs a single-parameter fit to $\stw$ with all other parameters
assumed to have their standard values, e.g., standard electroweak radiative
corrections with $\rho_0=1$.   
This fit determines
\begin{eqnarray}
    \sin^2\theta_W^{({\rms on-shell)}}&=&0.22773\pm0.00135({\rmt stat.})\pm0.00093({\rmt syst.})
        \nonumber\\
        &-&0.00022\times(\frac{M_{\rms top}^2-(175 \: \mathrm{GeV})^2}{(50 \: \mathrm{GeV})^2})
        \nonumber\\
        &+&0.00032\times \ln(\frac{M_{\rms Higgs}}{150 \: \mathrm{GeV}}).
\end{eqnarray}
The small dependences in $M_{\rms top}$ and $M_{\rms Higgs}$ result from
radiative corrections as determined from code supplied by
Bardin\cite{bardin} and from V6.34 of ZFITTER\cite{zfitter}; however, it
should be noted that these effects are small given existing constraints on
the top and Higgs masses\cite{LEPEWWG}.  A fit to the precision electroweak
data, excluding neutrino measurements, predicts a value of
$0.2227\pm0.00037$\cite{LEPEWWG,Martin}, approximately $3\sigma$ from the
NuTeV measurement.  Interpretations of the NuTeV data in terms of $M_W$ and
$\rho_\nu$ and model-independent neutrino-quark chiral couplings
are discussed elsewhere~\cite{nc-prl,ksm-lathuile}.

\section{Interpretation}

The NuTeV $\stw$ result is approximately three standard deviations
from the prediction of the standard electroweak theory.  This by
itself is surprising; however, it is not immediately apparent what the
cause of this discrepancy might be.  We discuss, in turn, the
possibility that the NuTeV result is a statistical fluctuation among
many precision results, the possibility that unexpected quark flavor
asymmetries or nuclear effects influence the result, and
finally possibilities for non-standard physics which could be
appearing in the anomalous NuTeV value.

\subsection{Significance in a Global Context}

For fits assuming the validity of the standard model, it is
appropriate to consider the {\em a priori} null hypothesis test chosen
in the proposal of the NuTeV experiment, namely the measurement of
$\stwos$.  The fit to precision data, including NuTeV, performed by
the LEPEWWG has a global the global $\chi^2$ of $28.8$ for $15$
degrees of freedom\cite{LEPEWWG,Martin}, 
including significant contributions from NuTeV's
$\stw$ measurement and $A_{FB}^{0,b}$ from LEP I.  The probability of the
fit $\chi^2$ being above $28.8$ is $1.7\%$.  Without NuTeV, this
probability of the resulting $\chi^2$ is a plausible $14\%$.  This
suggests that in the context of all the precision data, as compiled by
the LEPEWWG, the NuTeV result is still a statistical anomaly
sufficient to spoil, or at least sully, the fit within the standard model.

This large $\chi^2$ is dominated by two moderately discrepant
measurements, namely $A_{FB}^{0,b}$ and the NuTeV $\stw$, and if one
or both are discarded arbitrarily, then the data is reasonably
consistent with the standard model.  However, the procedure of merely
discarding one or both of these measurements to make the fit ``work''
is clearly not rigorous.  Furthermore, the potential danger of such a
procedure has been noted previously in the literature.  For example,
if $A_{FB}^{0,b}$ were disregarded, then the most favored value of the
Higgs mass from the fit would be well below the direct search limits.
Constraining the fit to be consistent with mass limits from
standard model Higgs boson searches results in still uncomfortably
large $\chi^2$\cite{Chanowitz}.

\subsection{Unexpected QCD Effects}

As noted above, corrections to Eqns.~\ref{eqn:ls} and \ref{eqn:rminus}
are required to extract electroweak parameters from neutrino
scattering on the NuTeV target.  In particular, these equations assume
targets symmetric under the exchange of $u$ and $d$ quarks, and that
quark seas consist of quarks and anti-quarks with identical momentum
distributions.

The NuTeV analysis corrects for the significant asymmetry of $d$ and $u$
quarks that arises because the NuTeV target, which is primarily composed of
iron, has an $\approx 6$\% fractional excess of neutrons over protons.
However, this correction is exact only with the assumption of isospin symmetry,
i.e., $\uubar_p(x)=\ddbar_n(x)$, $\ddbar_p(x)=\uubar_n(x)$.  This assumption,
if significantly incorrect, could produce a sizable effect in the NuTeV
extraction of $\stw$\cite{Sather,Thomas,Cao,Gambino}.

Dropping the assumptions of symmetric heavy quark seas, isospin symmetry and
a target symmetric in neutrons and protons, but assuming small deviations in
all cases, we calculate 
the effect of these deviations on $R^-$ is\cite{nc-asym}:
\begin{eqnarray}
\delta R^- & \approx 
& + \: \delta N \left( \frac{U_p-D_p}{U_p+D_p}\right) (3\Delta_u^2+\Delta_d^2) 
\nonumber\\
&& + \: \frac{(U_p-\Ubar_p-D_n+\Dbar_n)-(D_p-\Dbar_p-U_n+\Ubar_n)}{2(U_p-\Ubar_p+D_p-\Dbar_p)}  (3\Delta_u^2+\Delta_d^2)  \nonumber\\
&& + \: \frac{S_p-\Sbar_p}{U_p-\Ubar_p+D_p-\Dbar_p} (2\Delta_d^2-3(\Delta_d^2+\Delta_u^2)\epsilon_c),  
\label{eqn:deltaR-}
\end{eqnarray}
where $\Delta_{u,d}^2 = (\epsilon^{u,d}_L)^2-(\epsilon^{u,d}_R)^2$, $Q_N$ is
the total momentum carried by quark type $Q$ in nucleon $N$, and the neutron
excess, $\delta N \equiv A-2Z/A$.  $\epsilon_c$ denotes the ratio of the
scattering cross section from the strange sea including kinematic suppression
of heavy charm production to that without kinematic suppression.  The first
term is the effect of the neutron excess, which is accounted for in the NuTeV
analysis; the second is the effect of isospin violation and the third is the
effect of an asymmetric strange sea.

NuTeV does not exactly measure $R^-$, in part because it is not
possible experimentally to measure neutral current reactions down to
zero recoil energy.  To parameterize the exact effect of the symmetry
violations above, we have numerically evaluated the effects on the
NuTeV results of isospin and $s-\sbar$ asymmetries as a function of
$x$~\cite{nc-asym}.  This analysis shows that the level of isospin
violation required to shift the $\stw$ measured by NuTeV to its
standard model expectation would be, e.g., $D_p-U_n\sim0.01$ (about
5\% of $D_p+U_n$), and that the level of asymmetry in the strange sea
required would be $S-\Sbar\sim +0.007$ (about $30\%$ of $S+\Sbar$).

\subsubsection{Isospin Violations}

Several recent classes of non-perturbative models predict isospin
violation in the nucleon\cite{Sather,Thomas,Cao}.  The earliest
estimation in the literature, a bag model calculation\cite{Sather},
predicts large valence asymmetries of opposite sign in $u_p-d_n$ and
$d_p-u_n$ at all $x$, which would produce a shift in the NuTeV $\stw$
of $-0.0020$.  However, this estimate neglects a number of effects,
and a complete bag model calculation by Rodionov {\em et
al.\,}\cite{Thomas} conclude that asymmetries at very high $x$ are
larger, but the asymmetries at moderate $x$ are smaller and even of
opposite sign at low $x$, thereby reducing the shift in $\stw$ to a
negligible $-0.0001$.  Finally, the effect is also evaluated in the
meson cloud model\cite{Cao}, and there the asymmetries are much
smaller at all $x$, resulting in a modest shift in the NuTeV $\stw$ of
$+0.0002$.

Models aside, the NuTeV data itself cannot provide a significant independent
constraint on this form of isospin violation.  However, because PDFs
extracted from neutrino data (on heavy targets) are used to separate sea and
valence quark distributions which affect observables at hadron
colliders\cite{bodek}, global analyses of PDFs including the possibility of
isospin violation may be able to constrain this possibility
experimentally.  At least one author\cite{Kumano} has begun to consider the
experimental isospin constraints in the context of ``nuclear PDFs'',
and found very small isospin effects, except at very high $x$
and low $Q^2$, a region removed by the visible energy requirement
($E_{\rm calorimeter}>20$~GeV) of the NuTeV analysis.

\subsubsection{Strange Sea Asymmetry}

If the strange sea is generated by purely perturbative QCD processes, then
neglecting electromagnetic effects, one expects $\sav=\savbar$.  However, it
has been noted that non-perturbative QCD effects can generate a significant
momentum asymmetry between the strange and anti-strange
seas\cite{sNEsbar}.

By measuring the processes $\txnunub N\to \mu^+\mu^- X$ the CCFR and NuTeV
experiments constrain the difference between the momentum distributions of
the strange and anti-strange seas.
Within the NuTeV cross-section model, 
this data implies a {\em negative} 
asymmetry\cite{nc-asym},
\begin{equation}
 S-\Sbar = -0.0027 \pm 0.0013,
\end{equation}
\noindent
or an asymmetry of $11\pm6$\% of $(S+\Sbar)$.
Therefore, dropping the assumption of strange-antistrange symmetry results in
an {\em increase} in the NuTeV value of $\stw$,
\begin{equation}
 \Delta\stw = +0.0020 \pm 0.0009.
\end{equation}
\noindent
The initial NuTeV measurement, which assumes $\sav=\savbar$, becomes 
$$\stwos=0.2297\pm0.0019.$$ 
\noindent Hence, if we use the experimental measurement of the
strange sea asymmetry, the discrepancy with the standard model is increased
to $3.7\sigma$ significance.

\subsubsection{Nuclear Effects}

Nuclear effects which can be absorbed into process-independent PDFs
will not affect the NuTeV result.  However, several authors have
recently explored the possibility that neutrino neutral and charged
current reactions may see different nuclear effects and therefore
influence the NuTeV result.

A recent comment in the literature\cite{Thomas-Miller} has offered a
Vector Meson Dominance (VMD) model of low $x$ shadowing in which such an
effect might arise.  The most precise data that overlaps the low $x$
and $Q^2$ kinematic region of NuTeV comes from NMC\cite{NMC-arneodo},
which observed a logarithmic $Q^2$ dependence of the shadowing effect
as predicted by perturbative QCD for $Q^2$ independent shadowing as in
Pomeron models. However, models with a mixture of VMD and Pomeron
shadowing can be consistent with this high $Q^2$ data \cite{two-phase,
thomas-melnit-rant}.

The NuTeV analysis, which uses $\nu$ and $\nubar$ data at $<Q^2>$ of
$25$ and $16$~GeV$^2$, respectively, is far away from the VMD regime,
and the effect of this VMD model is significantly smaller than stated
in Ref.~\cite{Thomas-Miller}.  The most serious flaw in the hypothesis
that this accounts for the NuTeV result, however, is that it is not
internally consistent with the NuTeV data. Shadowing, a low $x$
phenomenon, largely affects the sea quark distributions which are
common between $\nu$ and $\nubar$ cross-sections, and therefore cancel
in $R^-$.  However, the effects in $\Rnu$ and $\Rnub$ individually
are much larger than in $R^-$ and this model {\em increases} the
prediction for NuTeV's $\Rnu$ and $\Rnub$ by $0.6\%$ and $1.2\%$,
respectively.  NuTeV's $\Rnu$ and $\Rnub$ are both below predictions
and the significant discrepancy is in the $\nu$ mode, not the $\nubar$
``control'' sample, both in serious contradiction with the prediction
of the VMD model.

Another recent paper\cite{Schmidt} has suggested that there may be
little or no EMC effect in the neutrino charged-current but the
expected EMC effect suppression at high $x$ in the neutral current.
If true, this could have the right behavior and perhaps magnitude to
explain the NuTeV data because of the effect at high $x$.
Unfortunately, this mechanism would cause large differences between
$F_2^\nu$ and $F_2^{\ell}$ on heavy targets at high $x$ which are
excluded by the CCFR charged-current cross-section
measurements\cite{unki-pmi}.

\subsection{New Physics}

The primary motivation for embarking on the NuTeV measurement was the
possibility of observing hints of new physics in a precise measurement
of neutrino-nucleon scattering.  NuTeV is well suited as a probe of
non-standard physics for two reasons. First, the precision of the
measurement is a significant improvement, most noticeably in
systematic uncertainties, over previous measurements. Second,
NuTeV's measurement has unique sensitivity to new processes when
compared to other precision data.  In particular, NuTeV probes weak
processes far off-shell, and thus is sensitive to other tree level
processes involving exchanges of heavy particles.  Also, the initial
state particle is a neutrino, and neutrino couplings are the most
poorly constrained by the $Z^0$ pole data, since they are primarily
accessed via the measurement of the $Z$ invisible width.

In considering models of new physics, a ``model-independent'' effective
coupling measurement~\cite{nc-prl,ksm-lathuile} is the best guide for
evaluating non-standard contributions to the NuTeV measurements.
This measurement suggests a large deviation in the left-handed chiral
coupling to the target quarks, while the right-handed coupling is as
expected.  Such a pattern of changes in couplings is consistent with
either a hypothesis of loop corrections that affect the weak process
itself or another tree level contribution that contributes primarily
to the left-handed coupling.  Chiral coupling deviations are often 
parameterized in
terms of the mass scale for a unit-coupling ``contact interaction'' in
analogy with the Fermi effective theory of low-energy weak
interactions.  Assuming a contact interaction described by a
Lagrangian of the form 
$$ 
-{\cal L}=\sum_{H_q\in\{L,R\}} 
\frac{\pm 4\pi} {\left( \Lambda^\pm_{LH_q}\right) ^2}\times
\left\{ \overline{l_L}\gamma^\mu l_{L}\overline{q_{H_q}}\gamma_\mu q_{H_q} 
  + l_{L}\gamma^\mu \overline{l_L}\overline{q_{H_q}}\gamma_\mu q_{H_q}\right.  
\left. + {\rmt C.C.}\right) ,
$$
the NuTeV result can be explained by an interaction with mass scale
$\Lambda_{LL}^+\approx 4\pm 0.8$~TeV.

\subsubsection{Extra U(1) Interaction}

Phenomenologically, an extra $U(1)$ gauge group which gives rise to
interactions mediated by a heavy $Z^{\prime}$ boson, $m_{Z^{\prime}}\gg m_Z$,
is an attractive model for new physics.  In general, the couplings associated
with this new interaction are arbitrary, although specific models in which a
new $U(1)$ arises may provide predictions or ranges of predictions for these
couplings.  An example of such a model is an $E(6)$ gauge group, which
encompasses the $SU(3)\times SU(2)\times U(1)$ of the standard model and also
predicts several additional $U(1)$ subgroups which lead to observable
interactions mediated by $Z^{\prime}$ bosons.
Before the NuTeV measurement, several authors had suggested in the literature
that the other precision electroweak data favored the possibility of a
$Z^{\prime}$ boson\cite{Zprime,Erler}.

We have analyzed the effect of $Z^{\prime}$s in $E(6)$ GUT
models\cite{langacker,E6-model} on the NuTeV measurement of the
chiral couplings. The effect of these bosons when the 
$Z$ and $Z^{\prime}$ do not mix is primarily on the
right-handed coupling.  It is possible to reduce the left-handed
coupling somewhat by allowing $Z-Z^{\prime}$ mixing; however, this
possibility is severely constrained by precision data at the $Z^0$
pole\cite{Erler}.

A $Z^{\prime}$ with coupling magnitudes equal to those of the $Z$
($Z^{\prime}_{SM}$) but leading to a destructive interference with the
$Z$ exchange could explain the NuTeV measurement if the $Z^{\prime}$
mass were in the range $\approx1$--$1.5$~TeV.  Current limits from the
TeVatron experiments on such $Z^{\prime}_{SM}$ are approximately
$0.7$~TeV\cite{TeV-zprime}.  Several authors have also
recently discussed other $U(1)$ extensions in the context of the NuTeV
result and found significant effects\cite{Gambino,Ma-Roy}.

\subsubsection{Anomalous Neutrino Neutral Current}

\begin{figure}
\begin{center}
\includegraphics[width=\textwidth]{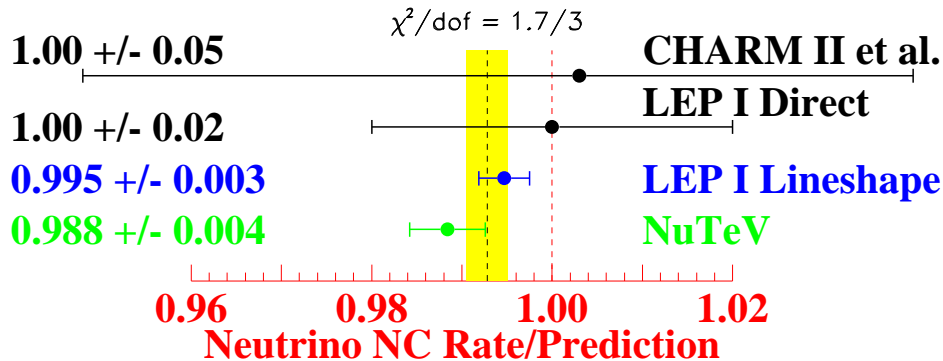}
\end{center}
\caption{\it Measurements of the neutrino current coupling, 
  interpreted as
  a neutrino neutral current interaction rate
  ($\propto\rho^{(\nu)}$).  The
  precise measurements, $\Gamma (Z\to\nu\nubar)$ at LEP~I and the NuTeV
  data, interpreted as an overall deviation in the strength of the neutral
  current coupling to neutrinos, are both below expectation.}
\label{fig:nunc}
\end{figure}

There are few precision measurements of neutrino neutral current
interactions.  Measurements of neutrino-electron scattering from the
CHARM~II experiment\cite{Charm2} and the direct measurement of
$\Gamma(Z\to\nu\nubar)$ from the observation of $Z\to\nu\nubar\gamma$
at the $Z^0$ pole\cite{LEPEWWG} provide measurements of a few percent
precision.  The two most precise measurements come from the inferred
$Z$ invisible width\cite{LEPEWWG} and NuTeV.  As is shown in
Figure~\ref{fig:nunc}, both of the precise rate measurements are
significantly below the expectation.  Theoretically, such a deviation
is difficult to accomodate.  One idea is a mixing of the light
neutrinos with another heavy gauge singlet, but this mechanism leads
to effects in {\em both} $Z\nu\nubar$ and $W\ell\nu$
vertices\cite{Gambino}.  However, Takeuchi and collaborators recently
suggested that both of these effects could be accomodated in the
precision electroweak data if the Higgs boson were
heavy\cite{Takeuchi}.

\section{Summary}

The NuTeV experiment has performed a measurement of $\stw$, and finds a
deviation of three standard deviations from the null hypothesis which assumes
the validity of the standard model of electroweak interactions.  Motivated by
the significance of this discrepancy, we study both conventional and new
physics explanations.  Several possibilities exist, although none is
theoretically compelling or has sufficient independent supporting evidence to
be a clear favorite.  Therefore, this result remains a puzzle.

\section*{Acknowledgements}
We gratefully acknowledge support for this work from the U.S.\
Department of Energy, the National Science Foundation and the Alfred
P.\ Sloan Foundation.  The NuTeV experiment benefitted greatly from
significant contributions from the Fermilab Particle Physics,
Computing, Technical and Beams Divisions.  In addition, we thank Stan
Brodsky, Jens Erler, Martin Gr\"{u}newald, Shunzo Kumano, Paul
Langackger, Jerry Miller, Michael Peskin, Jon Rosner, Ivan Schmidt and
Tony Thomas for useful input and discussions.


\begin{thebibliography}{99}
\def\np#1#2#3   {{ Nucl. Phys.} {\bf#1}, #2 (#3). }
\def\pcps#1#2#3 {{ Proc. Cam. Phil. Soc.} {\bf#1}, #2 (#3). }
\def\pl#1#2#3   {{ Phys. Lett.} {\bf#1}, #2 (#3). }
\def\plc#1#2#3   {{ Phys. Lett.} {\bf#1}, #2 (#3); }
\def\prep#1#2#3 {{ Phys. Rep.} {\bf#1}, #2 (#3). }
\def\prev#1#2#3 {{ Phys. Rev.} {\bf#1}, #2 (#3). }
\def\prl#1#2#3  {{ Phys. Rev. Lett.} {\bf#1}, #2 (#3). }
\def\prs#1#2#3  {{ Proc. Roy. Soc.} {\bf#1}, #2 (#3). }
\def\ptp#1#2#3  {{ Prog. Th. Phys.} {\bf#1}, #2 (#3). }
\def\rmp#1#2#3  {{ Rev. Mod. Phys.} {\bf#1}, #2 (#3). }
\def\rpp#1#2#3  {{ Rep. Prog. Phys.} {\bf#1}, #2 (#3). }
\def\zp#1#2#3   {{ Zeit. Phys.} {\bf#1}, #2 (#3). }
\def\epj#1#2#3   {{ Eur. Phys. Jour.} {\bf#1}, #2 (#3). }
\def\nim#1#2#3   {{ Nucl. Instr. Meth.} {\bf#1}, #2 (#3). }

\bibitem{langacker} P.~Langacker {\em et al.\,}, \rmp{64}{87}{1991}
\bibitem{oscpaper} K.~S.~McFarland, D.~Naples {\em et al.\,}, 
      \prl{75}{3993}{1995}
\bibitem{llewellyn} C.~H.~Llewellyn Smith, \np{B228}{205}{1983}
\bibitem{nc-prl} G.~P.~Zeller {\em et al.\,}, \prl{88}{091802}{2002}.
See also G.~P.~Zeller, Ph.D Thesis, Northwestern University (2002),
unpublished.
\bibitem{Paschos-Wolfenstein} E.~A.~Paschos and L.~Wolfenstein 
               \prev{D7}{91}{1973}
\bibitem{bardin} D.~Bardin and V.~A.~Dokuchaeva, JINR-E2-86-260 (1986).
\bibitem{zfitter} D.~Bardin {\em et al.}, Comp. Phys. Commun. 133 229 (2001).
\bibitem{LEPEWWG} 
CERN-EP/2001-98, hep-ex/0112021.
  Updated numbers used in this note are taken from
  http://lepewwg.web.cern.ch/LEPEWWG/ 
\bibitem{Martin} M.~Gr\"{u}newald,
  private communication, for the fit of Ref.~\cite{LEPEWWG} without
  neutrino-nucleon scattering data included.
\bibitem{ksm-lathuile} K.S.~McFarland {\em et al.\,}, 
hep-ex/0205080.
\bibitem{Chanowitz}M.~S.~Chanowitz, hep-ph/0207123.  M.~S.~Chanowitz,
\prl{87}{231802}{2001}. 
\bibitem{Sather} E.~Sather, \pl{B274}{433}{1992}
\bibitem{Thomas} E.~N.~Rodionov, A.~W.~Thomas, and J.~T.~Londergan,
Mod.\ Phys.\ Lett.\ A {\bf 9}, 1799 (1994).
\bibitem{Cao} F.~Cao and A.~I.~Signal, \prev{C62}{015203}{2000}
\bibitem{Gambino}
S.~Davidson, S.~Forte, P.~Gambino, N.~Rius, and A.~Strumia, hep-ph/0112302.
\bibitem{nc-asym} G.~P.~Zeller {\em et al.\,}, 
hep-ex/0203004.
\bibitem{bodek} A.~Bodek {\em et al.}, \prl{83}{2892}{1999}
\bibitem{Kumano} S.~Kumano, hep-ph/0209200.
\bibitem{sNEsbar} 
A.I.~Signal and A.W.~Thomas, \pl{B191}{205}{1987}.
M.~Burkardt and B.~J.~Warr, \prev{D45}{958}{1992}.
S.~Brodsky and B.~Ma, \pl{B381}{317}{1996}.
W.~Melnitchouk and M.~Malheiro, \pl{B451}{224}{1999}.
\bibitem{Thomas-Miller}
G.~A.~Miller and A.~W.~Thomas,
hep-ex/0204007.
\bibitem{NMC-arneodo}
M.~Arneodo {\em et al.\,} [NMC Collaboration],
\np{B481}{23}{1996}
\bibitem{two-phase} J.~Kwiecinski and B.~Badelek, Phys. Lett. {\bf B208}, 508
                  (1988).
\bibitem{thomas-melnit-rant} W.~Melnitchouk and A.~Thomas,
hep-ex/0208016.
\bibitem{Schmidt} S. Kovalenko et al., hep-ph/0207158. 
\bibitem{unki-pmi} U.~K.~Yang {\em et al.}, \prl{86}{2742}{2001}
\bibitem{Zprime} R.~Casalbuoni, S.~De Curtis, D.~Dominici 
and R.~Gatto, \pl{B460}{135}{1999}.
J.~L.~Rosner, \prev{D61}{016006}{2000}.
A.~Bodek and U.~Baur, \epj{C21}{607}{2001}.
\bibitem{Erler}
J.~Erler and P.~Langacker, \prl{84}{212}{2000}
\bibitem{E6-model}
G.~C.~Cho, K.~Hagiwara and Y.~Umeda, \np{B531}{65}{1998}.
D.~Zeppenfeld and K.~Cheung, hep-ph/9810277.
\bibitem{TeV-zprime}
F.~Abe {\it et al.}  [CDF Collaboration], \prl{79}{2192}{1997}.
B.~Abbott {\it et al.}  [D0 Collaboration], \prl{82}{4769}{2000}.
\bibitem{Ma-Roy} E.~Ma and D.~P.~Roy,
\prev{D65}{075021}{2002}
\bibitem{Charm2} P.~Vilain {\em et al\,}, Phys.\ Lett.\ {\bf B335} (1994) 248.
\bibitem{Takeuchi} T.~Takeuchi, hep-ph/0209109.
\end{thebibliography}
\end{document}